\documentclass[a4paper,12pt]{article}
\usepackage{jheppub,esint,shuffle,psfrag}
\usepackage[utf8]{inputenc}

\usepackage{epsfig,amssymb,amsmath,psfrag,subfigure,rotate,color,wasysym,xcolor}
\usepackage[bbgreekl]{mathbbol}

\allowdisplaybreaks
\newcommand{\insertfig}[2]{\includegraphics[width=#1cm]{#2}}

\DeclareSymbolFontAlphabet{\mathbbm}{bbold}
\DeclareSymbolFontAlphabet{\mathbb}{AMSb}%

\def\XXint#1#2#3{{\setbox0=\hbox{$#1{#2#3}{\int}$ }
\vcenter{\hbox{$#2#3$ }}\kern-.6\wd0}}

\def \be  {\begin{equation}}
\def \ee  {\end{equation}}
\def \ba  {\begin{eqnarray}}
\def \ea  {\end{eqnarray}}
\def \baa {\begin{eqnarray*}}
\def \eaa {\end{eqnarray*}}

\def \lab #1 {\label{#1}}

\newcommand\re[1]{(\ref{#1})}
\def\d{\hbox{{d}\kern-.20em\hbox{l}}}

\def \matrix #1 {\left(\begin{array}{cc} #1 \end{array}\right)}

\def \e  {\mathop{\rm e}\nolimits}

\newcommand \vev [1] {\langle{#1}\rangle}

\newcommand{\bit}[1]{\mbox{\boldmath$#1$}}
\def\1{\hbox{{1}\kern-.25em\hbox{l}}}

\newcommand{\ft}[2]{{\textstyle\frac{#1}{#2}}}

\makeatletter

\input pdf-trans
\newbox\qbox
\def\usecolor#1{\csname\string\color@#1\endcsname\space}
\newcommand\bordercolor[1]{\colsplit{1}{#1}}
\newcommand\fillcolor[1]{\colsplit{0}{#1}}
\newcommand\outline[1]{\leavevmode%
  \def\maltext{#1}%
  \setbox\qbox=\hbox{\maltext}%
  \boxgs{Q q 2 Tr \thickness\space w \fillcol\space \bordercol\space}{}%
  \copy\qbox%
}
\makeatother
\newcommand\colsplit[2]{\colorlet{tmpcolor}{#2}\edef\tmp{\usecolor{tmpcolor}}%
  \def\tmpB{}\expandafter\colsplithelp\tmp\relax%
  \ifnum0=#1\relax\edef\fillcol{\tmpB}\else\edef\bordercol{\tmpC}\fi}
\def\colsplithelp#1#2 #3\relax{%
  \edef\tmpB{\tmpB#1#2 }%
  \ifnum `#1>`9\relax\def\tmpC{#3}\else\colsplithelp#3\relax\fi
}
\bordercolor{black}
\fillcolor{white}
\def\thickness{.3}

\def\1{\mathbbm{1}}

\title{Near mass-shell double boxes}
 
\author[a]{A.V.~Belitsky,}
\author[b]{V.A. Smirnov}
\affiliation[a] {Department of Physics, Arizona State University,  Tempe, AZ 85287-1504, USA}  
\affiliation[b]{Skobeltsyn Institute of Nuclear Physics, Moscow State University 119992 Moscow, Russia}
\affiliation{Moscow Center for Fundamental and Applied Mathematics 119992 Moscow, Russia} 
  
 \abstract
{Two-loop multi-leg form factors in off-shell kinematics require knowledge of planar and nonplanar double box Feynman diagrams with 
massless internal propagators. These are complicated functions of Mandelstam variables and external particle virtualities. The latter serve 
as regulators of infrared divergences, thus making these observables finite in four space-time dimensions. In this paper, we use the method 
of canonical differential equations for calculation of (non)planar double box integrals in the near mass-shell kinematical regime, i.e., where 
virtualities of external particles are much smaller than the Mandelstam variables involved. We deduce a basis of master integrals with 
uniform transcendental weight based on the analysis of leading singularities by means of the Baikov representation as well as an array 
of complementary techniques. We dub the former asymptotically canonical since it is valid in the near mass-shell limit of interest. We 
iteratively solve resulting differential equations up to weight four in terms of multiple polylogarithms.
}

\begin{document}

\maketitle
\flushbottom
\setcounter{footnote} 0

\section{Introduction}
\label{s1}

Infrared structure of off-shell observables in massless gauge theories attracted attention in the past couple of years. Within the context 
of the maximally supersymmetric Yang-Mills theory (aka $\mathcal{N} = 4$ sYM) this kinematical regime can be addressed in a fully 
gauge-invariant fashion by studying the theory on its Coulomb branch \cite{HennGiggs1}. With a proper choice of vacuum expectation 
values for the model's scalar fields, one can mimic the off-shellness of the unbroken gauge symmetry phase with nonvanishing masses 
for external particles only, while keeping all states propagating in internal lines of Feynman graphs massless. This regime is of 
phenomenological interest in physical theories like QCD. As opposed to the fully massless case where infrared singularities arise as poles 
in the dimensional regularization parameter $\varepsilon = (4-D)/2$, in the nearly mass-shell regime of virtual amplitudes, they are replaced 
by the logarithms of external states' virtualities. An orthodox universality of infrared physics would suggest that critical exponents in both 
cases will be given by the very same function of the Yang-Mills coupling constant. However, recently this was demonstrated to be far from 
the truth.

Four- \cite{Caron-Huot:2021usw} and five-leg \cite{Bork:2022vat} scattering amplitudes as well as two-particle form factors 
\cite{Belitsky:2022itf,Belitsky:2023ssv} were the first few examples to exhibit a novel feature of the near mass-shell kinematics as 
opposed to the fully massless regime. While the infrared physics in the latter was known, since the inception of QCD 
\cite{Mueller:1981sg}, to be governed by the cusp anomalous dimension \cite{Polyakov:1980ca,Korchemsky:1987wg}  the former involved a 
different function of the coupling, the so-called octagon anomalous dimension \cite{Coronado:2018cxj,Belitsky:2019fan,Basso:2020xts}. 
To further elucidate its role, one needs to address more complicated observables containing more scales. They are of interest for several 
reasons. First, it is desirable to test the infrared factorization and off-shell universality in circumstances which involve multiple scales at 
the same time. Second, for near mass-shell scattering amplitudes with more than five legs and form factors with more than two, there is 
a residual finite contribution free from infrared logarithms but depending on Mandelstam-like variables. These are known as remainder 
functions. A natural question arises whether they are the same both in on- and off-shell regimes. Given that the critical exponents are 
different, one would expect them to differ as well. But one needs explicit verifications. 

Remainder functions in the massless case of planar $\mathcal{N} = 4$ sYM are endowed with a stringy description 
\cite{Alday:2007hr,Maldacena:2010kp} in terms of an effective two-dimensional world-sheet \cite{Basso:2013vsa,Sever:2020jjx}. The 
string in question is the so-called GKP string \cite{Gubser:2002tv} with its energy density determined by the cusp anomalous dimension. 
The string supports a set of elementary excitations of a T-dual theory, with their dispersion relations and scattering matrices known 
exactly in 't Hooft coupling constant \cite{Basso:2010in}. One may wonder then, given the vacuum of off-shell observables is determined 
by the octagon anomalous dimension, whether the spectrum of excitations that live on it and their interactions remain the same as on the 
GKP background. This is a long-term goal. On the way toward it, one needs `experimental' data from explicit field theoretical calculations 
to confirm or deny this expectation. A first step will be undertaken in this paper.

\begin{figure}[t]
\begin{center}
\mbox{
\begin{picture}(0,100)(220,0)
\put(0,-10){\insertfig{15}{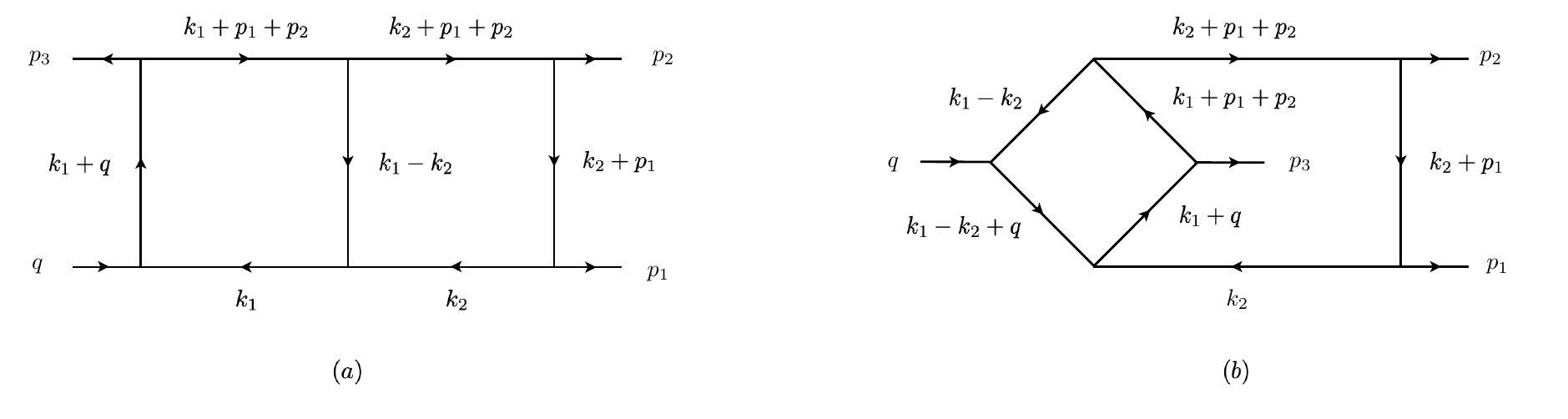}}
\end{picture}
}
\end{center}
\caption{\label{fig1} Planar and non-planar double box graphs in the left and right panels, respectively. }
\end{figure}

In this work, we calculate Feynman integrals of double box families, see Fig.\ \ref{fig1}, relevant for the problem of three-leg form 
factors $\vev{p_1,p_2,p_3|\mathcal{O} (q) |0}$ at two-loop perturbative order. As we advocated above, we are particularly interested 
in the kinematical situation where the off-shellnesses of external particles with momenta $p_\ell$ ($\ell =1,2,3$) are equal and small 
compared to the virtuality $q^2$ associated with the operator $\mathcal{O}$, --- the lowest super-component of the stress-tensor multiplet. 
To this end, we will rely on the method of differential equations \cite{Kotikov:1990kg,Gehrmann:1999as} in its modern day reincarnation 
that employs canonical bases \cite{Henn:2013pwa}. To date, this is the most powerful and efficient technique to tackle multi-scale 
Feynman integrals. Recently, it was successfully applied to the planar double box integrals (left panel in Fig.~\ref{fig1}) with three 
\cite{Dlapa:2021qsl} and four \cite{He:2022ctv} external squared momenta being off the light cone and all internal lines being massless. 
A basis of uniformly transcendental (UT) integrals was established and its symbol alphabet was analyzed. The latter was shown to be 
populated by letters expressed in terms of multiple square roots but no solution to the differential equations was offered. We will fill in this 
gap below for our kinematical situation and supplement it with an analysis of nonplanar graphs as well (right panel in Fig.~\ref{fig1}) which 
are far more complex. We will demonstrate that the near mass-shell limit provides sufficient simplification of the structure of canonical 
differential equations to offer a solution in terms of multiple polylogarithms \cite{Goncharov:2001iea} with all square roots gone from their 
arguments.

Our subsequent presentation is organized as follows. In the next section, we set up the kinematics and classes of Feynman integrals to 
be studied both for planar and nonplanar families. Next, we move on to the construction of canonical bases. We will discuss the two cases 
in parallel providing necessary technical details as needed so that a curious reader could reproduce all results if desired. We start in 
Section \ref{SectionPreliminaryMIs} by building an initial basis of master integrals (MIs) making use of the integration-by-parts (IBP) to deduce 
a primary basis. Then in Section \ref{SectionCanonicalBasis}, we use leading singularities and a variety of other available techniques 
to cast them in the canonical form for the planar case. The nonplanar family if far more complex and while we manage to build canonical 
basis for lower sectors, we encounter elliptic cases and thus turn to the asymptotic limit in question where these degenerate into poles. The 
asymptotically canonical basis for non-planar graphs is presented in Section \ref{SectionAsyCanonicalBasis}. We provide solutions to the 
resulting differential equations and determine corresponding integration constants in Section \ref{SectionIntegration} making use of a variety 
of criteria which bypass the necessity for explicit evaluation of parametric integrals. In both situations, explicit results are given up to weight-four 
in terms of multiple polylogarithms and they are further recast in terms of classical Euler polylogarithms and an additional two-argument 
polylogarithm ${\rm Li}_{2,2}$ at weight-four. Mathematica notebooks and ancillary files attached with this submission spill sufficient `secrets 
of the trade' for a newbie to familiarize oneself with the subject and can be regarded as `blueprints' for construction and solution of canonical 
differential equations for any (non-elliptic) family of Feynman graphs.

\section{Setting up conventions}

To begin with, let us establish our notations. The kinematics of interest corresponds to the momentum flow from the operator source 
$\mathcal{O} (q)$ to off-shell external particles, obeying the conservation condition $q = p_1 + p_2 + p_3$. We introduce three 
Mandelstam variables according to
\begin{align}
u \equiv - (p_1 + p_2)^2
\, , \qquad
v \equiv - (p_2 + p_3)^2
\, , \qquad
w \equiv - (p_3 + p_1)^2
\, , 
\end{align}
which are related by the equation
\begin{align}
u+v+w = - q^2 + 3 \mu
\, .
\end{align}
Nevertheless, throughout our subsequent analysis, we will treat $u$, $v$ and $w$ as independent since this provides stringent
checks on the correctness of our derivations. Above, we introduced equal Euclidean virtualities for all particle legs $p_\ell^2 \equiv - 
\mu$. Since the overall scale of a Feynman integral can be always unambiguously restored on dimensional grounds, we will set 
$q^2 = - 1$ in what follows. 

The families of graphs which take centre stage in this paper are shown in Fig.\ \ref{fig1}. In spite of the fact that Feynman integrals
contributing to physical observables are finite for nonvanishing off-shellness, nevertheless we will work with a dimensionally regularized
theory in order to be able to apply IBP reduction which requires a $D$-dimensional setup to render bases of sought after MIs complete. 
The two-loop non- and planar integrals 
\begin{align}
G_{a_1 \dots a_9} 
= 
\e^{2 \varepsilon \gamma_{\rm E}}
\int \frac{d^D k_1}{i \pi^{D/2}} \frac{d^D k_2}{i \pi^{D/2}} \prod_{j=1}^9 D_j^{-a_j}
\end{align}
are determined by a set of massless propagators $1/D_j$ (for $j = 2, \dots, 7$) and two irreducible scalar products $D_{8,9}$ defined
according to the momentum flow exhibited in Fig.\ \ref{fig1} as
\begin{align}
&D_2 = - (k_1 + p_1 + p_2)^2
\, ,
& 
&D_3 = - (k_1 + q)^2
\, ,
&
&D_4 = - (k_1 - k_2)^2
\, ,
& 
&D_5 = - k_2^2
\, , \\
&D_6 = - (k_2 + p_1)^2
\, , 
&
&D_7 = - (k_2 + p_1 + p_2)^2
\, ,
& 
&D_8 = - (k_1 + p_1)^2
\, ,
& 
&D_9 = - (k_2 + q)^2
\, , \nonumber
\end{align}
with the remaining denominator
\begin{equation}
D_1 = - k_1^2
\, , \qquad\mbox{and}\qquad
D_1 = - (k_1 - k_2 + q)^2
\end{equation}
corresponding to the planar and non-planar cases, respectively. All indices $a_i$ are integers with $a_{8,9} \leq 0$. Let us turn to these 
two families one by one.

\section{Primary basis of MIs and differential equations}
\label{SectionPreliminaryMIs}

Let us begin with the planar graph as a case study. It was previously addressed in Refs.\ \cite{Dlapa:2021qsl,He:2022ctv}, however,
we will use this more familiar family to set up our formalism so that we could be more concise in our following presentation of the 
non-planar graph, which is computationally more demanding but does not bring anything new to the table to a certain degree.

Preliminary counting of MIs can be done with a variety of available tools, say with {\tt Mint} \cite{Lee:2013hzt} or the modular component 
of {\tt FIRE} \cite{Smirnov:2023yhb}, which was the go-to tool in our analysis. Constructing a list of integrals in the top, i.e., level-seven 
sector, obtained by inequivalent permutations of two indices equal to two $G_{221111100}$, we prepare start files and generate 
symmetry relations with {\tt LiteRed} \cite{Lee:2012cn,Lee:2013mka}. A modular arithmetic IBP then yields an initial set of 74 MIs. We give 
preference to Laporta-reducible values of indices being equal to 2 since experience with canonical bases have taught us that these more likely 
than not be endowed with single leading singularities and thus serve as UT candidates. Next, we use {\tt FindRules} command of {\tt FIRE} 
to deduce 10 symmetry equations between MIs in our preferential basis, thus reducing their number to 64. However, this is not the end of 
the story. We can further determine `hidden' relations as well. To accomplish this, we create lists of integrals sufficiently close to the preliminary 
set and containing these as a subset: it includes integrals with none, one and two indices set to 2. Then an IBP reduction reveals additional 
two relations among them reducing their total number to 62. At this step, it is always advisable to verify that thus obtained basis does not yield 
the so-called `bad' denominators according to the nomenclature of Ref.\ \cite{Smirnov:2020quc}. In fact, we find none. But at level five, IBP 
yields quite lengthy denominator polynomials in the Mandelstam variables and virtuality and these can be traded however for a significantly 
more compact ones. This provides us with a solid starting set of 62 preliminary MIs $\bit{I}$ for our subsequent analytical analysis which we 
will use from now on as the option {\tt \#masters} for IBP reduction with {\tt FIRE}. All of the above steps are presented in Sections 1 through 
10 of the attached Mathematica notebook {\tt A2Zdbox.nb} and output is saved in the subdirectory {\tt dbox}. An analysis identical to the one 
just discussed is performed then for the non-planar family to give us a set of 97 MIs $\bit{I}$ and stored in the subdirectory {\tt nbox}.

After these preparatory studies, we move on to the construction of the derivatives in the Mandelstam variables and virtuality by
performing differentiations with {\tt LiteRed}. The differential $d \bit{I} = du \, \partial_u \bit{I} +  dv \, \partial_v \bit{I} +  dw \, \partial_w \bit{I} 
+  d\mu \, \partial_\mu \bit{I}$ then needs to be IBP-reduced back to the MIs $\bit{I}$ thus generating the sought-after differential equations
\begin{align}
\label{IniDEs}
\partial_i \bit{I} = \bit{M}_i \cdot \bit{I}
\, ,
\end{align}
with $i = u,v,w,\mu$. While the analytic IBP reduction for the planar family takes a matter of hours on a typical machine, the non-planar case 
is far more computationally demanding. For instance, the reduction of level-seven integrals from the left-hand side of the differential equations 
\re{IniDEs} given in the accompanying file {\tt intsde-nbox2.m} down to MIs \#3 $G_{010110000}$ and \#4 $G_{101001000}$ 
from {\tt pr-nbox2.m} takes a staggering 7 and 10 days, respectively. To cross-check that the resulting tables are indeed correct, we relied on 
modular arithmetic runs with {\tt FIRE}  with subsequent balanced rational reconstruction developed in Ref.\ \cite{Belitsky:2023qho}. With an 
MPI parallelization of 1024 cores of ASU's {\tt Sol} supercluster \cite{SolSC}, the $7 \to \#3$ IBP took 58 hours with {\tt Flint} \cite{Flint} and 
46 hours with {\tt Symbolica} \cite{Symbolica} but indeed confirmed our earlier analytical findings. We provide detailed account of the derivation 
in Section 11 of the accompanying notebook {\tt A2Zdbox.nb} for the planar graph.

\section{Canonical basis}
\label{SectionCanonicalBasis}

Now the main task at hand is to transform the basis of MIs $\bit{I} = \bit{T} \cdot \bit{J}$ such that the differential equations \re{IniDEs} admit 
their canonical form 
\begin{align}
\label{CanDEs}
\partial_i \bit{J} = \varepsilon \bit{A}_i \cdot \bit{J}
\, , \qquad
\varepsilon \bit{A}_i = \bit{T}^{-1} \cdot \bit{M} \cdot \bit{T} - \bit{T}^{-1} \cdot \partial_i \bit{T}
\, , 
\end{align}
with each element of the $\bit{A}$-matrices being Fuchsian, i.e., possessing simple poles only, and $\varepsilon$-independent \cite{Henn:2013pwa}. 
To this end, we need to determine viable UT candidates from our primary list of MIs. Provided this procedure is successful, a differentiation of these 
pure UT integrals will reduce their transcendental weight by one and, thus, the right-hand side of \re{IniDEs} will have to be proportional to 
$\varepsilon$, which carries the transcendentality weight $-1$. To practically implement this strategy, we rely on the well-known conjecture that 
connects uniform weight integrals with the properties of their integrands, namely, that the singularities of an integrand are locally of logarithmic 
type \cite{Arkani-Hamed:2012zlh,Bern:2014kca}, see, e.g., \cite{Wasser:2022kwg} for a comprehensive review. 

As the calculation of unitarity cuts is in general downright easier than solution of integrals per se, the idea is to use the former for the 
identification of Feynman integrals that correspond to pure functions. To perform multidimensional unitarity cuts efficiently, one has to rely on an 
appropriate parametrization. Since Feynman integrals possess integrands which are rational functions of propagators and ISPs, it is only natural 
to choose these as integration parameters $z_i \equiv D_i$. To date this is the most concise framework which is known as the Baikov representation 
\cite{Baikov:1996rk,Baikov:2005nv}, see Refs.\ \cite{Dlapa:2022nct,Bosma:2022xkn} for comprehensive reviews. This form of integrals trivializes 
computation of unitarity cuts. The so-called leading singularities correspond to taking the maximal cut, i.e., successive residues in all $z_i = 0$, 
followed by residues in composite singularities emerging along the way from any Jacobian factors \cite{Arkani-Hamed:2010pyv}. This completely 
localizes all integrations and provides a function of external kinematical variables, which once being divided out from the Feynman integral in 
question yields a pure UT candidate with constant leading singularity. Of course, for a given integral, there could be multiple ways to localize all 
integrations depending on the order of taking residues and this can yield different leading singularities. Only integrals with a single leading singularity 
can be autonomously recast as UT, while in cases where there are more that one, linear combinations of these have to be studied as well. It is 
important to realize that UT candidates found this way may not correspond to MIs of the traditional Laporta algorithm. This is the reason why we 
chose from the very beginning to favor MIs having indices equal to 2 in our IBP studies. Leading singularities analysis completely fixes the
diagonal blocks of the $\bit{A}$-matrices, which do not mix MIs at different levels. Then we move on to study sub-maximal cuts to find
corrections from lower-level subsectors.

In our analysis, we relied on the Mathematica implementation of the Baikov parametrization via {\tt Baikov.m} package developed in Ref.\ 
\cite{Frellesvig:2017aai}. Starting from the lowest sector, we iteratively constructed the canonical basis of MIs for the planar graph. Since 
a more general case was already studied in the literature \cite{He:2022ctv}, we provide only sporadic details in Sections 12-14 of the ancillary 
file {\tt A2Zdbox.nb} with final results for the canonical basis and all $\bit{A}$-matrices given in {\tt dboxCan62.m} and {\tt AuPC.m}, 
{\tt AvPC.m}, {\tt AwPC.m}, {\tt AmPC.m}, respectively.

\section{Asymptotically canonical basis}
\label{SectionAsyCanonicalBasis}

The analysis alluded to above immediately convinces us that the non-planar graph (right panel in Fig.\ \ref{fig1}) possesses elliptic sectors 
\cite{Broedel:2018qkq}, see Ref.\ \cite{Weinzierl:2022eaz} for a thorough review, implying that some leading singularities reside on elliptic 
curves \cite{Bourjaily:2020hjv,Bourjaily:2021vyj} rather being merely algebraic. However, they smoothly degenerate into the latter as we send 
the virtuality $\mu$ down to zero. Since at the end of the day, all we are after is the asymptotic behavior of our MIs as $\mu \to 0$ up to terms 
which vanish in $\mu$, we can implement this limit on differential equations for the primary set of 97 MIs and then construct what we call as 
the {\sl asymptotically} canonical basis.

Thus, we change the strategy for basis construction in the non-planar case. Namely, we tend to assume generic values for all variables
($u$, $v$, $w$ as well as $\mu$) as long as we encounter only algebraic leading singularities and swiftly pass to the asymptotic consideration 
when it is no longer the case. In particular, this occurs in the two level-six sectors $G_{111101100}$ and $G_{111111000}$, 
and the top level-seven sector $G_{111111100}$. For these, we require the following properties to be fulfilled by the differential
equations: (i) the `virtuality' matrix $\bit{M}_{\mu}$ can be cast in the form
\begin{equation}
\label{AsyCanAmu}
\bit{M}_{\mu} \to \bit{A}_\mu = \frac{\varepsilon}{\mu} \bit{A}^0_\mu+O(\mu^0)
\, ,
\end{equation}
with $\bit{A}^0_\mu$ being a matrix of rational numbers, i.e., its elements are strictly independent of  $u$, $v$ and $w$; (ii) $\bit{M}_i$ matrices 
for the Mandelstam variables $i = (u,v,w)$ have well-defined finite limit as $\mu \to 0$ and do not possess elements with square roots; (iii) last but 
not least, resulting differential equations in $u$, $v$ and $w$ are canonical or can be made canonical with an appropriate similarity transformation.

As usual, we focus on diagonal blocks first but use now the leading order form of the elements of the matrices $\bit{M}_i$ (with $i =u,v,w,\mu$) 
as $\mu$ goes to zero for a proper choice of asymptotically canonical elements. It is easier to demonstrate it with a example of the
$G_{111101100}$ sector. The primary set of MIs defining this sector is
\begin{align}
\{
G_{111101100}
\, , \
G_{111101200}
\, , \
&
G_{111102100}
\, , \
G_{111201100}
\, ,  \\
&
G_{112101100}
\, , \
G_{121101100}
\}
\, . \nonumber
\end{align}
To start with we utilize the form of $\bit{M}_\mu = \bit{M}^0_\mu/\mu + O (\mu)$, with $\bit{M}^0_\mu$ being a function of $u,v,w$ and 
$\varepsilon$, to conclude that if we are to multiply the elements 2,3,5,6 with $\mu$, after a similarity transformation the virtuality matrix will 
take the required form \re{AsyCanAmu}. We know a priori, however, that the $\varepsilon$-dependence of this seed basis will have to be adjusted 
later since, as rule of thumb, one associates $\varepsilon^4$ with MIs without any twos and $\varepsilon^{2-n}$ for MIs with $n$ twos in the 
first 7 positions. For now, it will do the job however. Next, we change the basis by multiplying each element in it with an unknown function of the 
Mandelstam variables
\begin{align}
G \to f (u,v,w) G
\, .
\end{align}
Enforcing the $\varepsilon$-form on the resulting differential equations for this new basis, we solve the arising differential equations on
the functions $f (u,v,w)$ and get, after a gentle mixture of elements with each other and re-arrangement,
\begin{align}
\label{NaiveLevel6nbox}
\big\{
&
\mu v (u+v+w)^2/(u+v) G_{111101200} \, , \
\mu v [ w G_{111102100} + (u+v+w) G_{111101200} ] \, , \nonumber\\
&
\mu v [ u G_{112101100} + (u+v+w) G_{121101100} ] \, , \
\mu v (u+v+w)^2/(v+w) G_{121101100} \, ,  \nonumber\\
&\qquad\qquad\qquad\qquad\qquad\qquad\qquad\qquad\qquad\qquad
f_5 G_{111101100} \, , \
f_6 G_{111201100}
\big\}
\, .
\end{align}
The last two elements of this naive basis contain square roots of Mandelstam variables in functions $f_{5,6}$. This is hardly a surprise
since we completely ignored up to now the correct $\varepsilon$-dependence of the basis elements which resulted in erroneous
differential equations for $f_{5,6}$. If we do it in a proper manner, we observe a violation of the canonical form of the differential equations 
in $u$, $v$ and $w$. Details on this calculation can be found in Section 1 of the accompanying Mathematica notebook {\tt AsyClevel6.nb}.

The above finding instructs us to look further for a better choice of the elements 5 and 6 of this sector. The method of trial and error is quite 
tedious and exhausting, so we turn to the massless non-planar double box for inspiration. In the strict limit $\mu \to 0$, the first four elements 
if \re{NaiveLevel6nbox} vanish and one is left with just two elements. The massless non-planar box analysis demonstrates that indeed 
it possesses two level-six sectors with one of them containing two primary elements $G_{111101100}$ and 
$G_{111101200}$. Construction of the canonical basis in this sector is performed in Section 2 of the notebook {\tt AsyClevel6.nb} 
and offers two options for UT elements, namely, 
\begin{align}
\label{Option1Massless}
\{ v (u + v + w) G_{111101100} \, , \ (v + w) G_{1111011-10} \}
\, ,
\end{align}
or
\begin{align}
\label{Option2Massless}
\{ v (u + v + w) G_{111101100} \, , \  (u + v) G_{11110110-1} \}
\, .
\end{align}
We then build upon \re{Option1Massless} to lift the analysis to the off-shell case in the asymptotic limit and construct the final form
of the diagonal block of this sector in Section 3 of {\tt AsyClevel6.nb} such that the last two elements in Eq.\ \re{NaiveLevel6nbox} are
replaced with 
\begin{align}
&
(1 + 4 \varepsilon) \varepsilon v (u+v+w) G_{111101100} \\
&\qquad
+
\varepsilon \mu v  (u+v+w)/(u+v) [ w G_{111102100} - (u+v+w) G_{111101200}] \nonumber\\
&\qquad
+
\varepsilon \mu v (u+v+w)/(v+w) [u G_{112101100} - (u+v+w) G_{121101100}] \nonumber
\end{align}
and 
\begin{align}
&
(1 + 4 \varepsilon) \varepsilon (v+w) G_{1111011-10} \\
&\quad
+
\ft14 \varepsilon \mu v /(v+w) [ u (2 u + 5 v+ 5 w)G_{112101100} - (u+v+w) (2 u+v+w) G_{121101100} ]
\, , \nonumber
\end{align}
respectively, and we simultaneously restored a proper relative $\varepsilon$-normalization of these elements. All other sectors can
be analyzed in the same fashion.

Though all diagonal sectors were brought by us to the $\varepsilon$-form, not all elements of the $\bit{A}$-matrices are Fuchsian. 
Moreover the off-diagonal blocks are not even close to the required $\varepsilon$-form. However, their transformation to the canonical 
form is now purely algorithmic. The Fuchsian form is easily obtained making use of the code {\tt Canonica.m} 
\cite{Meyer:2016slj,Meyer:2017joq,Meyer:2018feh}. The latter cannot handle, however, transformation of all off-diagonal elements in the 
differential equations to the $\varepsilon$-form using the Lee's trick \cite{Lee:2014ioa}. The latter is implemented in a powerful package 
{\tt Libra.m} \cite{Lee:2020zfb}, which calls for an external {\tt Fermat} \cite{Fermat} computer algebra system, though it works just as 
fine\footnote{Though ten times slower.} even with built-in Mathematica commands. This systematic procedure is demonstrated step-by-step  
in the ancillary notebook {\tt AsyCnbox.nb} proving the output in the file {\tt nboxAsyCan97.m} in the {\tt Canonica}  format. This culminates
our quest to reach the asymptotically canonical basis in the non-planar case.

\section{Integration and determination of integration constants}
\label{SectionIntegration}

To summarize, in the previous section we determined the asymptotically canonical form of the differential equation with the one in the
off-shellness $\mu$ being 
\begin{equation}
\label{DEs0}
\partial_\mu \bit{J} =\frac{\varepsilon}{\mu} \bit{A}^0_\mu  \bit{J}
\, ,
\end{equation}
up to terms vanishing as $\mu \to 0$, with $\bit{A}^0_\mu$ being a purely numerical matrix. Solving this leading order equation is simple 
and it provides a transformation
\begin{equation}
\bit{J} (u,v,w,\mu) =  \mu^{\varepsilon \bit{\scriptstyle A}^0_\mu} \cdot \bit{J}_0 (u,v,w) 
\, ,
\end{equation}
to the $\mu$-independent basis $\bit{J}_0$ via a matrix exponent whose entries are expressed as linear combinations of 
$\mu^{- n \varepsilon}$-terms accompanied by rational number coefficients. The basis $\bit{J}_0$ solves in turn the asymptotically 
canonical equations 
\begin{equation}
\label{ZeroMuDEs}
\partial_i \bit{J}_0 = \varepsilon  \bit{A}^0_i  \bit{J}_0
\qquad\mbox{with}\qquad i = (u,v,w)
\, ,
\end{equation}
where the elements of $\bit{A}^0_i = \bit{A}_i |_{\mu = 0}$ are expressed in terms of rational functions of the Mandelstam invariants 
$u$, $v$ and $w$ only.  

The main advantage of the canonical form of differential equations \re{ZeroMuDEs} is that their solution can be written in terms of a 
path-ordered exponential and explicitly calculated via Chen's iterative integrals theory \cite{Chen1977IteratedPI},
\begin{align}
\label{PathIntegralSolution}
\bit{J}_0
=
P_\gamma \exp \left( \varepsilon \int_\gamma \bit{A}^0 \right) \bit{J}_{00}
\, ,
\end{align}
where $\bit{A}^0 = du \bit{A}^0_u + dv \bit{A}^0_v + dw \bit{A}^0_w$ and $\bit{J}_{00}$ is a vector of integration constants. At each order in the 
$\varepsilon$-expansion it receives an independent set of unknowns
\begin{align}
\bit{J}_{00} = \sum_{p \geq 0} \varepsilon^p \bit{c}^{(p)} 
\, .
\end{align}
The solution \re{PathIntegralSolution} is independent of the choice of the path $\gamma$ since the integrability of the differential equations 
is a zero-curvature condition, ${\rm d}  \bit{A}^0 - \varepsilon \bit{A}^0 \wedge \bit{A}^0 = 0$. In our analysis, we chose a piece-wise path
\begin{align}
\gamma = [0,u] \cup [0,v] \cup [0,w]
\, .
\end{align}
In more practical terms, we take the first equation with $i = u$ in (\ref{ZeroMuDEs}) and solve it with respect to $u$. Then we turn to the
$v$-variable. To eliminate the $u$-dependence from the differential equation, we form a difference between the right-hand side of 
(\ref{ZeroMuDEs}) for $i = v$ and the derivative in $v$ of the solution found at the previous step. We then find its primitive in $v$. Finally,
we repeat the procedure for $w$. This procedure is cast in a Mathematica module {\tt Integrator} in the attached notebook  {\tt AsySolCnbox.nb}.
The result is then given in terms of multiple polylogarithms (MPLs) \cite{Goncharov:2001iea}. The latter are defined 
recursively via an integral iteration, e.g., 
\begin{align*}
G(a_0, \bit{a};u) = \int_{[0,u]}du' \frac{G(\bit{a};u')}{u' - a_0} 
\, . 
\end{align*}
However, in order to have a better handle on the analytical structure of our results, we recast them in terms of classical Euler 
polylogarithms\footnote{These are readily encoded in Mathematica.} whenever possible. It is well known that for the weight up to (and including) 
three, all MPLs can be traded to ${\rm Li}_n$'s, see, e.g., \cite{MR618278,Zhao2001MotivicCO,ZhaoBook} (Chapter 2 on MPLs of the latter 
reference is available at \cite{ZhaoChapter2}). At weight four, one needs to include an additional two variable MPL ${\rm Li}_{2,2}$ to the minimal 
basis of classical polylogarithms as was observed in Ref.\ \cite{Duhr:2011zq}. This basis transformation was worked out and is conveniently 
implemented into the routine {\tt gtolrules.m} devised in Ref.\ \cite{Frellesvig:2016ske}. We just need to make sure that a proper `branch' of 
MPLs is taken into account at a given point in the $(u,v,w)$-space since a single expression in terms of ${\rm Li}_n$'s and ${\rm Li}_{2,2}$ is 
not sufficient to cover the entire space of (complexified) Mandelstam variables.

At each $p$-order of the $\varepsilon$-expansion, arrays of integration constants $\bit{c}^{(p)}$ have to be determined from a set of 
boundary conditions. However, we would like to avoid an explicit calculation of any Feynman integrals since, even in some corners of the 
phase space, they are very complex and, which is worse, quite numerous. Instead, we relied on several criteria to fix them such as (i) 
numerology, (ii) cancellation of unphysical poles, (iii) absence of imaginary parts and (iv) finite integrals. 

(i) The first condition is self-explanatory. Using the fact that our asymptotically canonical MIs obey the property of being UT,  
we cast the integration constants into a product of rational numbers times values of the Riemann zeta function $\zeta_p = \zeta (p)$ of the same 
transcendental weight,
\begin{align}
\bit{c}^{(p)} = \bit{r}^{(p)} \zeta_p
\, ,
\end{align}
with the employed convention $\zeta_0 = 1$ and $\zeta_1 = 0$ for the first two values of $p$. Then, computing the MIs at a random point
for the Mandelstam variables with FIESTA \cite{Smirnov:2021rhf}, we confronted its results against numerical evaluation of our solutions.
In this manner, we managed to confidently determine 84 $r^{(p)}_j$ for $p=0,2$, 81 $r^{(3)}_j$'s and 79 $r^{(4)}_j$'s. The monotonically
decreasing number of rationally reconstructed constants with increasing $p$ is related to the loss of FIESTA's numerical precision and 
emergence of large rationals at higher orders in $\varepsilon$.

(ii) To alleviate the aforementioned problem and cross check correctness of previous numerological findings, we employed conditions for 
unphysical pole cancellation in the right-hand side of the canonical differential equations, namely, at $u+v = 0$, $v+w = 0$ and $w+u = 0$. 
Then, decomposing the $A$-matrices in explicitly Fuchsian form
\begin{align}
\bit{A}^0_i = \frac{\bit{a}_{i,u+v}}{u + v} + \frac{\bit{a}_{i,v+w}}{v + w} + \frac{\bit{a}_{i,w+u}}{w + u} + \dots
\, ,
\end{align}
we imposed the following equations on our basis
\begin{align}
\label{NoUnphysicalPoles}
\bit{a}_{i,\alpha} \bit{J}_0|_{\alpha = 0} = 0
\, .
\end{align}
These provided a further set of 10,7,9 and 10 identifications/relations between integration constants at levels $p=0,2,3$ and $4$, respectively.

(iii) To further constrain the integration constants at order $p$, we used solutions at order $p+1$ and required vanishing of imaginary parts
as one approaches unphysical poles in Eqs.\ \re{NoUnphysicalPoles}. This provided the value on the last $r^{(3)}_{90}$ element from the 
solution at order $\varepsilon^4$. The solution at fifth order in $\varepsilon$ was used in conjunction with high-precision numerical 
computations of MPLs with the C++ package {\tt GiNaC} \cite{Bauer:2000cp} making use of a Mathematica interface from Ref.\ \cite{Duhr:2019tlz} 
and subsequent reconstruction of analytical expressions with the PSLQ algorithm \cite{PSLQ:1999}. This allowed us to deduce 3 equations 
for $r^{(4)}_j$ ($j=86,92,97$). 

(iv) The implementation of the above three conditions fixed all but 3,6,7 and 8 integration constants for $p=0,2,3$ and $4$, correspondingly. 
Then, by a judicious choice, we found a set of 26 finite (in $\varepsilon$) integrals
\begin{align}
& 
G_{011111100} \, , 
&&
G_{101111100} \, ,  
&&
G_{110111100} \, ,
&&
G_{111011100} \, ,   
&&
G_{111110100} \, , \nonumber\\  
& 
G_{111111000} \, , 
&&
G_{1111111-10} \, , 
&&
G_{11111110-1} \, , 
&&
G_{111111100} \, , 
&&
G_{1111111-1-1} \, , \nonumber\\ 
& G_{011101100} \, , 
&&
G_{011110100} \, ,  
&&
G_{011111000} \, , 
&&
G_{101101100} \, , 
&&
G_{101110100} \, ,  \nonumber\\
&  
G_{101111000} \, , 
&&
G_{110101100} \, , 
&&
G_{110111000} \, ,  
&&
G_{111001100} \, ,  
&&
G_{111011000} \, ,  \nonumber\\
&  
G_{111100100} \, , 
&&
G_{111101000} \, , 
&&
G_{1111011-10} \, ,
&&
G_{111101100} \, ,
&&
G_{1111110-10} \, , \nonumber\\
&
&&
&&
G_{11111100-1} \, ,
&&
\end{align}
which were reduced with IBP identities to our set of 97 MIs. The resulting relations are divergent and pole cancellation in the Laurent 
$\varepsilon$-expansion provided an ultimate set of equations to completely fix the solutions at orders one through three. At fourth order,
we obtained the last five integration constants whose numerical value to $O (10^{-3})$ were determined to be
\begin{align}
r^{(4)}_{83} = 1515.669 \, , \
r^{(4)}_{84} = 26.958 \, ,  \
r^{(4)}_{90} = -50645.784 \, , \
r^{(4)}_{91} = 6.659 \, , \
r^{(4)}_{95} = -576.338
\, .
\end{align}
To rationalize these, one has to either perform an analytic calculation of a very large set of Feynman integrals or increase the accuracy of 
their numerical evaluation to twelve decimal places with FIESTA or any other program. Currently, alas, this is beyond our reach. However,
a particular combination of these constants shows up in the three-leg form factor \cite{ToBePublished}, which allows us to eliminate one of 
them from the above list. It reads
\begin{align}
\frac{r^{(4)}_{83}}{42}-r^{(4)}_{84}+\frac{r^{(4)}_{90}}{252}-\frac{r^{(4)}_{91}}{6}+\frac{r^{(4)}_{95}}{8}
= -
\frac{62683849}{236544}
\, .
\end{align}

All steps of the above analysis for the non-planar family are thoroughly presented in the accompanying Mathematica notebook 
{\tt AsySolCnbox.nb}. For the planar graph, it suffices to use just the first two conditions (i) and (ii). All solutions up to the same order 
in $\varepsilon$ are quoted in the ancillary file {\tt AsySolCdbox.nb}.
 
\section{Conclusions}

With this paper, we initiate a series of studies of multi-scale two-loop observables in $\mathcal{N} = 4$ sYM. Currently, we constructed 
bases of UT MIs for double box planar and non-planar graphs in the kinematical limit of small 
virtualities of three external particles and an arbitrary invariant mass for the last leg. This is a preparatory study for a full-fledged analysis 
of the three-leg form factor of the stress tensor multiplet to be published separately \cite{ToBePublished}. The bases in question were used to 
determine canonical form of differential equations in the Mandelstam variables $u$,  $v$,  $w$ as well as the off-shellness $\mu$. Solutions 
to these equations were constructed up to terms vanishing as $\mu \to 0$. The results for all master integrals were obtained as a double 
Laurent/Taylor expansion in $\varepsilon$/$\log \mu$ up to (and including) weight four contributions. All integration constants were successfully
fixed analytically for the exception of 5 coefficients at level four where they were found numerically with the accuracy $10^{-3}$. Future
more precise numerical studies with FIESTA could potentially fix them unambiguously as rational coefficients accompanying the value of 
$\zeta_4$.

Consideration performed in this work will be generalized in a number of avenues. From the point of view of identifying two-dimensional 
integrable physics of the octagon flux-tube, form factors of super-descendants of the stress-tensor multiplet could provide simpler 
circumstances for its elucidation since they are sensitive to contribution from single charged flux-tube excitations \cite{Basso:2023bwv} 
as opposed to singlet pairs determining the lowest half-BPS operator \cite{Sever:2020jjx}. So far the limitation to form factor observables
was solely driven by a lower-multiplicity requirement on the number of external legs in a graph to attain a non-trivial remainder function.
It is well-known that in the case of scattering amplitudes, nontrivial remainder functions emerge starting from six legs. Thus, it is important 
to analyze these in the near mass-shell kinematics introduced in this paper and compare them both functionally as well as from the 
microscopic stringy point of view.

Regarding development of computational techniques of multi-loop Feynman integrals per se, we are currently capable to break free from 
the simplifying assumption of the near mass-shell limit and uplift our asymptotically canonical basis for the non-planar graph to the situation 
of arbitrary virtualities. Solution to the resulting equations is a very different issue though.

\begin{acknowledgments}
We are deeply indebted to Roman Lee and Alexander Smirnov for their generous help at various stages of this project. The work of A.B.\ was 
supported by the U.S.\ National Science Foundation under the grant No.\ PHY-2207138. The work of V.S. was supported by the Russian Science 
Foundation under the agreement No.\ 21-71-30003 (IBP reduction with rational reconstruction) and by the Ministry of Education and Science of 
the Russian Federation as part of the program of the Moscow Center for Fundamental and Applied Mathematics under Agreement No.\ 
075-15-2022-284 (solving differential equations with asymptotically canonical bases).
\end{acknowledgments}


\end{document}